# Quantitative and Computational Radiobiology for Precision Radiopharmaceutical Therapies

**Authors and Affiliations:**

**Tahir Yusufaly, PhD (Corresponding Author)**

Assistant Professor, Department of Radiology, Johns Hopkins, Baltimore, MD

**Hamid Abdollahi, PhD**

Research Associate, Department of Basic and Translational Research, BC Cancer Research Institute, Vancouver, Canada

**Babak Saboury, MD, MPH**
Radiology and Imaging Sciences, Clinical Center, National Institutes of Health, Bethesda, MD

**Arman Rahmim, PhD, DABSNM**
Professor, Department of Radiology, Physics and Biomedical Engineering, University of British Columbia, Vancouver, Canada; Distinguished Scientist, Integrative Oncology, BC Cancer Research Institute, Vancouver, Canada

**Key Points:**

- Quantitative and computational radiobiology enables the prediction of clinical outcomes and treatment effectiveness from physical dosimetry and patient-specific anatomy and physiology.
- The standard linear-quadratic (LQ) paradigm of radiobiology is rooted in external beam radiation therapy (EBRT), and has served dosimetry and treatment planning in that modality quite well.
- The radiobiology of radiopharmaceutical therapy (RPT) differs from that of EBRT, due to the idiosyncrasies of pharmacokinetically-mediated dosimetry that modulate radiation response in ways that are not fully understood.
- First-principles multiscale simulations, including single-cell track structure and multicellular agent-based calculations, serve as virtual quantitative microscopes that can help guide us in the development of a complementary radiobiological paradigm for RPT.
- Progress in RPT radiobiology for dosimetry-driven treatment planning will require a combination of mechanistic and phenomenological modeling, *in vitro* experiments, and clinical translation.

## I.     Introduction

Radiation biology (hereafter radiobiology), the science of the effects of ionizing radiation on living organisms, is an indispensable component in the toolkit of modern-day radiation and nuclear medicine (*1*). Radiobiology forms the bridge linking physical dosimetry to observed health outcomes and clinical endpoints, and is the scientific basis behind why certain organs have higher or lower tolerance doses than others, or why some cancers require more or less coverage for disease control.

Radiation response is a complex process that depends on a wide panoply of physical and biological factors. Compared to most other physical and chemical agents, our collective knowledge of the mechanisms underlying response to ionizing radiation is relatively advanced and, moreover, fairly quantitative (*2,3*). A repertoire of mathematical and computational methods exist for describing the etiological mechanisms and phenomenological consequences of radiation exposure. These advances have mostly been focused on radiobiology in relation to external beam radiation therapy (EBRT).

However, the radiobiology of radiopharmaceutical therapies (RPTs) is a much younger area of study, and has rapidly evolved into a dynamic and interdisciplinary field (*4–6*). RPT response is shaped by an intricate interplay of physical and biological, pharmacological, and immunological factors. The emergence of diverse agents spanning emitters including α-, β-, and Auger-emitters has introduced unique spatiotemporal dose distributions that modulate radiation response in unknown ways.

These recent advances have highlighted that there are still significant gaps in our knowledge of the underlying mechanisms of radiation response, and have also forced the community to collectively re-examine mathematical modeling paradigms of EBRT radiobiology as applied to RPT. They have raised critical questions regarding the extent to which the linear-quadratic (LQ) formalism (*7*) used in EBRT accurately reflects the biology of radiopharmaceutical exposures, and how to extend or go beyond it in situations where this is not the case. These issues must be grappled with before we can clearly determine what constitutes a biological effective dose (BED (*8*)) in RPT, and whether current clinical trial endpoints are an adequate representation of the full spectrum of RPT effects.

Such knowledge gaps are not merely academic; they carry significant clinical consequences. Without a nuanced understanding of how to quantitatively predict radiobiological responses, we risk misestimating treatment outcomes, missing opportunities for personalized therapies, and overlooking synergistic potentials in combination therapies (*9*). As the field advances, it will thus be imperative to revisit foundational assumptions in radiobiological modeling. To this end, `first-principles' multiscale simulations of radiation response can play an important role (*2,10*). These simulations are built around the philosophy of modelling, in as much detail and realism as possible, the underlying mechanistic processes in radiobiology, including subcellular damage as well as intracellular and multicellular response.

A skeptical reader might ask if such a multiscale approach is truly necessary – after all, one could just take the image-estimated tumor and organ absorbed doses and assume that efficacy and toxicity are simple logistic functions of these inputs, with phenomenological parameters that can be determined empirically. Over the short-term, such an approach will indeed likely be the most realistic starting point in the practical development of dosimetry-driven treatment planning approaches for RPT. However, over the long-term, we believe that a more nuanced multiscale understanding of biological effectiveness, and how it is modified by physical and patient-specific factors, will be necessary to realize the full potential of RPT for patient outcomes.

With this background in mind, this review proceeds to give an overview of quantitative and computational radiobiology for RPT. We start in section II by briefly reviewing the standard LQ paradigm used in modern EBRT clinics. In section III, we introduce a deeper dive into the underlying mechanisms and present an overview of the current state-of-the-art in first principles multiscale computations of radiation response. In section IV, we discuss how these calculations can be used to gain insight and address radiobiological knowledge gaps that arise due to the idiosyncrasies of RPT. Section V concludes the article with an outlook and some speculative thoughts.

## II. Clinical EBRT Radiobiology in the LQ Paradigm

**Figure 1** displays a graphical summary of the basic elements of the clinical LQ radiobiology paradigm as it is commonly used in EBRT. The starting point in this paradigm (*7*) is the calculation of cell survival using the formula

$$SF = e^{-(\alpha D + \beta G D^2)} \text{ or } \ln SF = -\alpha D - \beta G D^2 \quad (1)$$

Here, SF is survival fraction, $\alpha$ and $\beta$ are the radiosensitivity parameters of the cells, D is the absorbed dose to the cell, and G is the dose-rate factor, which for an acutely delivered single fraction is equal to 1. The $\alpha$ parameter represents 'one-hit' directly lethal damage from a single radiation track, while the $\beta$ parameter represents 'multiple-hit' accumulated sublethal damage (SLD) due to multiple tracks. The curvature of the LQ dose-response curve is characterized by the $\alpha/\beta$ ratio, with units of Gy, which corresponds to dose at which the linear and quadratic contributions are equal.

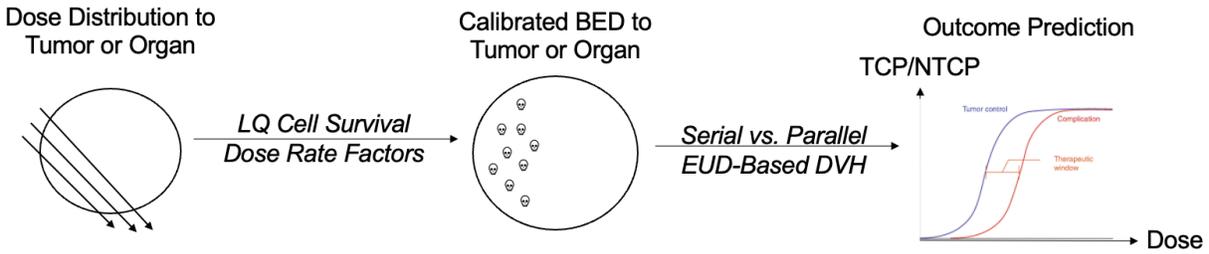

**Figure 1:** Schematic overview of the end-to-end LQ paradigm for clinical EBRT radiobiology, including calculation of the biological effective dose with dose rate factors, followed by estimation of an equivalent uniform dose / dose volume histogram metric that has a sigmoidal association with tumor control (blue) and/or normal tissue complication (red), from which one can define the concept of a therapeutic window (red).

For protracted or fractionated dose rates, equation 1 must also include calculation of the G-factor to account for the kinetics of sublethal damage (SLD) repair. Lea and Catcheside (*11*) showed that for mono-exponential repair kinetics, with a time-dependent dose rate $\dot{D}(t)$, the G factor can be derived to be:

$$G(T, \mu_R) = \frac{2}{D^2} \int_0^T \dot{D}(t) dt \int_0^t \dot{D}(w) e^{-\mu_R(t-w)} dw \quad (2)$$

where T is the total duration of the treatment and $\mu_R$ is the SLD repair constant. This cell survival model can be used to derive a formula for the BED (*8*,*12*):

$$BED = D \left(1 + \frac{G(T, \mu_R) D}{\alpha/\beta}\right) \quad (3)$$

This formula for the BED can also be extended to include other complicating factors, such as tumor regrowth or reoxygenation (*13*), although it should be noted that these are not as widely used clinically.

After the estimation of the BED across different targets, the final step is to translate this information into aggregate clinically observable outcomes, namely tumor control probability (TCP

(*14*)) and normal tissue complication probability (NTCP (*15*)). Both of these are influenced by the spatial nonuniformity of the BED and corresponding cell death probability. In addition, it is important to remember that the form of the TCP or NTCP curve depends on the choice of the clinical endpoint. The TCP for progression-free survival, in general, will depend on different tumor dosimetric variables than those of disease-free survival. Likewise, for NTCP, toxicity endpoints can be either acute or chronic, and a careful specification of the timescale of measurement and the choice of endpoints is crucial for a robust and reliable calibration of NTCP.

Equivalent uniform dose (EUD) is a concept introduced (*16,17*) to account for spatial heterogeneities. The EUD of a given nonuniform dose distribution is formally defined as the same dose that, if delivered uniformly, will yield the same TCP (for tumors) or NTCP (for normal organs). EUD was originally suggested for tumors by Niemierko, and was then generalized to also apply to normal organs, resulting in the latter sometimes being referred to as the generalized equivalent uniform dose (gEUD). For convenience, we will use EUD from hereon. Formally, the EUD is defined as

$$EUD = \left(\frac{1}{N}\sum_{i=1}^{N} d_i^a\right)^{1/a} \quad (4)$$

Here, N is the number of tissue or tumor voxels, d is the dose in a voxel (we implicitly assume here that it has been calibrated to an appropriate BED), and a is the volume exponent, which is negative for tumors and positive for normal organs. For a = 1, the EUD equals the mean dose $D_{mean}$, typical of normal organs with a 'parallel' functional organization (e.g., bone marrow). For *a* >> 1, the EUD approaches the maximum dose $D_{max}$. more typical of normal organs with a 'serial' functional organization (e.g., spinal cords) where the dose hot spots are the bottleneck for overall organ toxicity. For *a* << 0, EUD approaches the minimum dose $D_{min}$, a situation more associated with tumors, where the underdosed regions of the tumor are the ones likely to prevent tumor control. More generally, the EUD for varying values of a can be interpreted as a tunable dose-volume histogram (DVH) metric. This EUD or DVH metric is ultimately mapped to TCP or NTCP via a sigmoid-like function, a common choice being the Lyman Kutcher Burman (LKB) formula (*18,19*):

$$TCP \text{ or } NTCP = \frac{1}{\sqrt{2\pi}} \int_{-\infty}^{t} e^{-\frac{x^2}{2}} dx \text{ with } t = \frac{EUD - TD_{50,5}}{m \, TD_{50,5}} \quad (5)$$

Here, *m* is the steepness of the dose–response curve, and TD50,5 is the EUD value for which 50% of the population exhibited complications within 5 years for a uniform whole-organ or whole-tumor irradiation.

### III. First-Principles Multiscale Radiobiology: State-of-the-Art

The traditional LQ radiobiological paradigm has served radiation oncology well, and is the scientific workhorse underlying modern treatment planning and optimization in EBRT. However, as mentioned in the introduction, radiopharmaceutical exposures interact with cells and tissues via a different set of mechanisms than external exposures, which makes the black-box application of EBRT absorbed dose and DVH paradigms in general inadequate for the prediction of therapeutic efficacy and toxicity. In our opinion, without a more rigorous multiscale understanding of how dose is linked to molecular events and systemic response, and how these vary for different RPT exposures, RPT cannot hope to achieve its full potential. To this end, in this section we introduce a deeper dive into the mechanisms underlying radiation response, and the state-of-the-art tools for modeling them computationally. **Figure 2** displays a graphical summary of these methods in an end-to-end format analogous to Figure 1 for the clinical LQ paradigm.

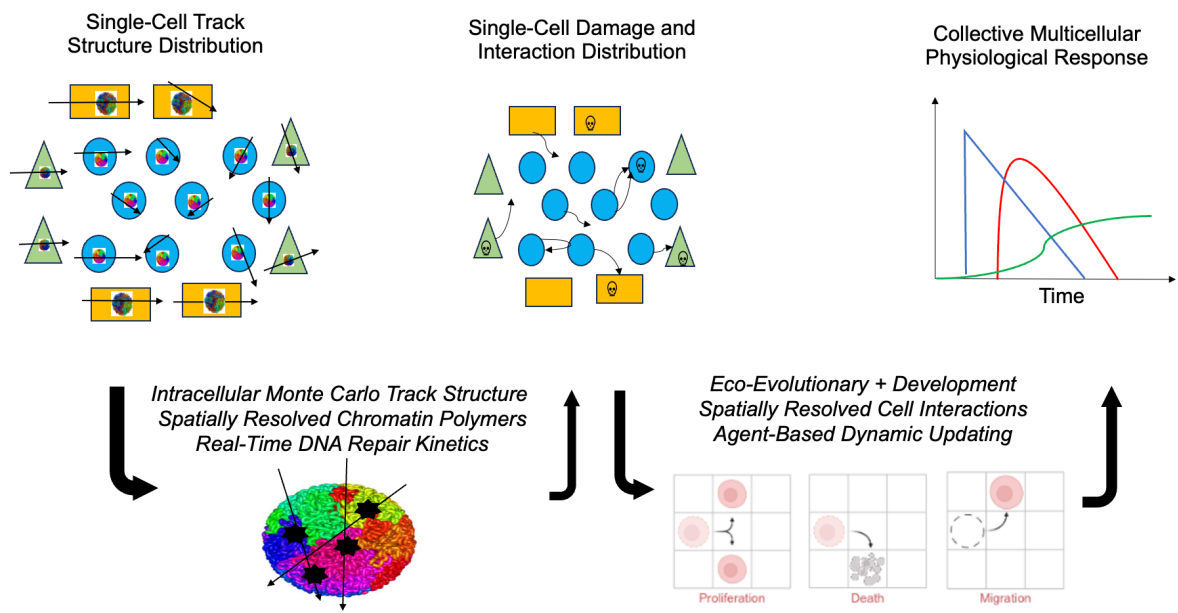

**Figure 2:** Schematic overview of the current state of the art in end-to-end first principles multiscale radiobiological calculations. The universal starting point for these calculations is the distribution of track structures across cells. These are paired with Monte Carlo molecular modeling of the intracellular polymer physics and repair kinetics. The results of these microsimulations then feed into higher level agent-based models for simulating the collective multicellular response of an organ or tumor to the induction of injury.

The first step in the radiation response process is the induction of DNA damage across the cell nucleus from the incident radiation track structure. This step is, by far, the one that has been most thoroughly studied computationally (*20,21*). Our collective ability to quantitatively predict DNA damage, while imperfect, is vastly superior to our ability to predict in detail anything that happens after, whether intracellular (repair kinetics, cell cycle exit) or multicellular (signaling, regeneration, and tumor evolution).

While it is clear that DNA damage alone is insufficient for characterizing the full spectrum of radiobiological response, its importance should still not be understated. In particular, when considering questions of how radiobiology in RPT differs from EBRT, track structures are really the only truly fundamental 'common denominator'. These calculations, therefore, are the most robust and general strategy for calibrating radiobiological modeling across the two modalities. Any radiobiological modeling paradigm that starts downstream of track structure necessarily involves approximations and coarse-graining of the details of radiation dose and subcellular biology. Track structures are the only starting point that is truly *ab initio* across the spectrum of possible patterns of radiation delivery.

Radiation track structure calculations, using molecular modeling and Monte Carlo simulations, explicitly account for the interactions of individual charged particles (electrons, protons, ions, etc.) with the atoms and molecules of the intracellular medium, in particular the cell nucleus. These include both direct interactions, where the charged particles themselves modify the DNA, and indirect interactions, where the charged particles ionize free radical species that mediate the resulting damage. The different radiation chemistry processes, such as ionization, excitation, or elastic scattering, have been well characterized at the atomic level for calculations in a uniform water medium. When coupled with polymer physics models (*22*) of the dynamic chromatin fiber, these allow for the estimation of spatially resolved maps of DNA lesions at chromosomal, and even base-pair, resolution. The most important of these lesions that drive radiobiological response are double strand breaks (DSBs) and, to a lesser extent, single strand breaks (SSBs). However, other types of lesions, such as single-base modifications or mismatches, can also play a role, in particular by influencing the complexity of DSBs and SSBs, which in turn influences DNA damage response (23).

Myriad tools exist for calculating the initial biological damage, including Geant4-DNA (24) with the TOPAS-nBIO platform (25), Monte Carlo Damage Simulation (MCDS (26)), and recent GPU-accelerated packages (27,28). More recently, these tools have started expanding to incorporate modeling of processes downstream of DNA damage. This includes both: 1) processes at the single-cell level, including intracellular repair kinetics and cell fate determination, and 2) processes at the multicellular level, including the ecology and evolution of tumor microenvironments and the pathophysiological dynamics of wound healing and regeneration in normal organs.

At the intracellular level, the dominant radiobiological response is driven by DNA repair kinetics, in particular for DSBs. The three most prominent repair pathways for radiation-induced DNA DSBs are homologous recombination (HR), non-homologous end rejoining (NHEJ) and microhomology mediated end joining (MMEJ). The key molecular players and events of these pathways have been fairly well studied, although of course open questions still remain. These include considerations such as how the choice of DSB repair is influenced by cell cycle phase, the interaction of DSB repair with other competing damage response pathways, and the influence of structural and chemical complexities in the DSB and intracellular environment. Nevertheless, the key qualitative mechanisms of how DNA misrepair, lethal aberrations, and early and late cell cycle exit couple are mostly in place by now. Our collective understanding of these mechanisms has sufficiently advanced to allow for the development of *in silico* extensions to TOPAS-nBIO, in particular the MEDRAS (Mechanistic DNA Repair and Survival (29)) and DaMaRis (DNA Mechanistic Repair Simulator (30)) models.

Beyond single-cell damage, however, the complete characterization of dynamic response across an aggregate tumor or organ must also consider larger-scale multicellular processes. Physiological wound-healing is characterized by a complex coordinated sequence of events (31,32), including intercellular signaling, inflammation, stem cell proliferation and differentiation, and fibrotic scarring. At diseased sites, these steps are modulated by, and in turn cause change in, the heterogeneous tumor microenvironment. This results in a complex series of feedback loops that drive tumor ecology and evolution (33–35), many of the details of which remain only partially understood.

Advances in the larger computational biology and oncology fields have led to the emergence of agent-based simulation (36) as a dominant paradigm for modeling multicellular systems. While such calculations come in many flavors, at a broad general level they are unified in starting with a heterogeneous population of cells across a spatial environment. Based on the possible states of these cells, the modeler can define possible dynamic transition events that a cell can undergo, such as proliferation, differentiation, or migration. The probability of a given cellular agent undergoing any specific transitions is determined by a prespecified set of rules that depend on both intrinsic characteristics of the cells themselves, as well as any interactions they may have with each other or with their microenvironment. The choice of possible states, transitions, or interaction rules to include in the simulation is flexible, and can be adapted as needed based on knowledge of the underlying intercellular organization and histological interactions.

In recent years, tools in the radiation research community have emerged adapting agent-based modeling methods for radiobiology. Notably, the CompuCell3D platform, widely used for agent-based modeling of multicellular ecology, evolution, and development, has been coupled to state-of-the-art radiobiological simulation. Examples include the RADCELL module (*37*) bridging Geant4-DNA with CompuCell3D and the TOPAS-tissue framework (*38*) for extending TOPAS-nBIO.

## IV. Towards a Clinical Radiobiology for RPT: when should we build on the LQ paradigm and when should we create new paradigms *de novo*?

As our discussion in section III has highlighted, our knowledge of the mechanisms of radiation response and how to mathematically model them is relatively sophisticated compared to most other agents. Nevertheless, the sheer complexity of the processes involved reinforces the point that significant uncertainties remain regarding many of the details. Moreover, given the intrinsically dynamic and diverse nature of biological systems, it is arguable that these uncertainties will never be completely eliminated.

Given these challenges, it is in many ways nothing short of extraordinary that the LQ paradigm for EBRT radiobiology works at all, let alone that it works as well as it does. The key equations introduced in section II predate our knowledge of many of the finer details of radiation response. Historically, these formulas were justified on the basis of heuristic arguments, such as the combination of single-hit and multi-hit lesions (for the LQ formula) or the serial and parallel dichotomy of volume effects (for the EUD). While these justifications may at one point have been considered etiologically sound, as our knowledge of the details has accumulated it has become clearer that it is more appropriate to think of LQ as an `effective' phenomenological model.

Moreover, it has also become apparent that we do not understand when, how and why this effective simple description emerges. In turn, this incomplete understanding has resulted in us being blind to the exact scope of applicability of the traditional paradigm. We understand that a given set of LQ parameters or a specific formula for the BED should not be expected to transfer to any arbitrary situation. Likewise, we recognize that while EUD and DVH-based metrics are convenient and useful, they remain surrogates to more fundamental underlying functional organization. And we intuitively understand that the more 'EBRT-like' the radiation exposure, the more likely that the LQ paradigm will give reliable results. However, we cannot in general anticipate exactly how much an exposure needs to deviate from EBRT-typical conditions before traditional paradigms and models become unreliable.

When considering how to extrapolate to the RPT setting, such ignorance becomes especially problematic. Internal radionuclide dosimetry is fundamentally different from external dosimetry by virtue of being mediated (*39–41*) by patient-specific pharmacokinetics (PK). In RPT, the heterogeneity of target expression, binding and internalization dynamics, and macro-to-micro circulation of the agent in the blood all impact the actual dose delivered to tissues. Such levels of biological dependency are far less important in physical dose calculation in EBRT. In addition, even when the actual physical dose rate and radiation quality to cells is similar, single-cell BEDs may be further modified via additional extraneous modifying factors in the local chemical environment. For example, in somatostatin antagonist-based theranostics (*42*) the differential binding and internalization of antagonists compared to agonists can modulate intracellular response via a series of indirect and incompletely understood pathways.

Thus, it is only through an understanding of PK can we come to appreciate the differences between EBRT and RPT exposures that, in turn, drive differences in radiobiological response. At a general level, the differences group into two categories:

1. 'Local' differences in single-cell exposures: In EBRT, neighboring cells receive essentially identical, coherent and uniform quality dose and dose rates. In RPT, the combination of cell-specific uptake and finite range of the emitters results in different cell populations receiving a more stochastic distribution of doses with a heterogeneous spectrum of dose rates and radiation qualities (*43*).

2. 'Global' differences across the multicellular structure: In EBRT, the large-scale dose gradients are driven by tumor anatomy and treatment planning, such that the dose that a given cell receives is determined by its location with respect to the tumor. In RPT, different cell types co-existing in the same local vicinity can result in different statistical distributions of radiation, and it is the spatial localization of the RPT agent that drives large-scale dose gradients rather than tumor anatomy.

In recent years, several groups (*44–47*) have started investigating local differences through mechanistic modeling of cellular and subcellular track structure. These studies have highlighted the importance of microdosimetric fluctuations, variability in relative biological effectiveness, dose rate effects, and interactions with chemical and immunological agents as all being factors that modify overall cell survival. Other schools of research (*48*) have looked into whether some of these modifying factors can be incorporated into the conventional LQ modeling framework via appropriate extensions.

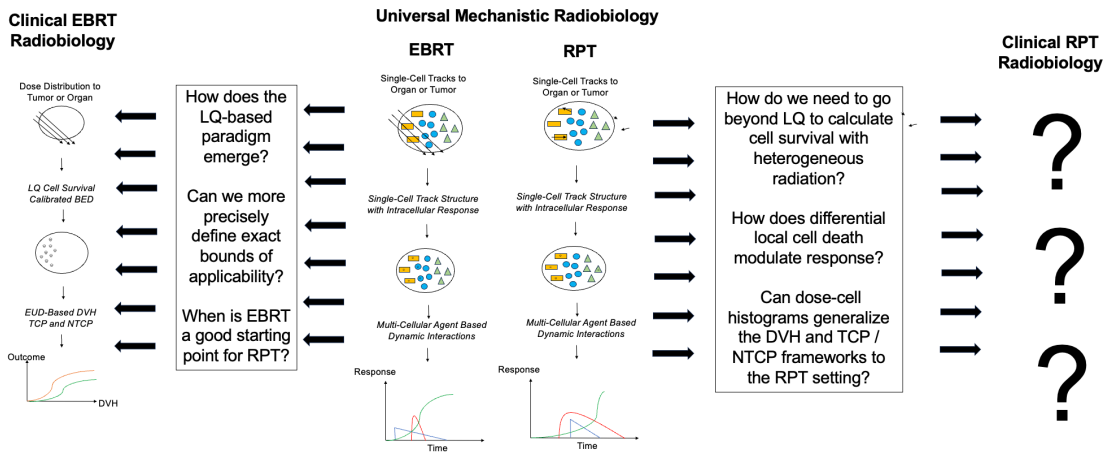

**Figure 3:** Given the complexity of the underlying universal mechanistic radiobiology, it is in many ways remarkable that clinical EBRT radiobiology with the LQ-based BED paradigm (left) works as well as it does. Moving forward, our increased knowledge of the underlying mechanisms (center) motivates us to try to understand the reasons for this unreasonable effectiveness. Such an understanding, in turn, will have direct implications in our search for a corresponding parsimonious RPT radiobiology that can be used practically in the clinic (right).

Moving forward, a fruitful direction in our opinion would be to see if these two strands of thought can be unified into an integrated multiscale framework, such that we understand exactly how phenomenological LQ factors emerge from judicious approximations to the underlying microscopic complexity. Moreover, such efforts will also help identify situations where local differences between RPT and EBRT are so great that building on the LQ is simply infeasible, indicating a need for entirely new cell survival models altogether. A schematic illustration of this proposal is shown in **Figure 3**.

However, while there has been progress on the local front, studies investigating differences in the global multicellular response between EBRT and RPT are comparatively sparse. A critical challenge for such modeling has been the difficulty in acquiring experimental information and guidance on small scale dosimetry (*49,50*), which is shaped by the histological organizations of different cells and cell types. This organization constrains the possible patterns of biodistribution and uptake of RPT agents. This, in turn, constrains possible dose-cell histograms (DCHs), which may be thought of as generalizations of whole organ and tumor DVHs that distinguish between the heterogeneous cell types in the tissue population.

Thus, understanding the differences in multicellular response between EBRT and RPT fundamentally comes down to understanding how the response to a multivariate spectrum of DCHs differs from that of a univariate DVH. A noteworthy example of the importance of understanding this difference is the recent evidence (*51*) that alpha-emitters display an enhanced abscopal effect, with elevated TCP at absorbed doses well below expected EBRT thresholds. Although the reasons for these and related effects remain unclear, it is suggested that differential sparing of local immune cells in the tumor microenvironment, due to heterogeneous uptake and short particle range, plays an essential role. Much more agent-based modeling and experimental validation is needed to investigate the collective dynamic response to heterogeneous cell death distributions, in a way that goes beyond specific clinical endpoints or approximate surrogates and starts from the truly fundamental processes underlying their emergence, as highlighted in **Figure 3**.

## V. Conclusions

Quantitative and computational radiobiology provides a powerful framework for linking physical dosimetry and patient-specific factors to biological outcomes in RPT, enabling a deeper and more mechanistically grounded understanding of therapeutic effects. By spanning the full continuum—from pharmacokinetics and image-based dose estimation to cell survival modeling and prediction of TCP and NTCP —this approach can help translate absorbed dose into clinically actionable insights. Traditional models and paradigms from EBRT such as the LQ and EUD formalisms remain foundational, but must be adapted to capture the unique spatiotemporal and biological complexities of modern RPT. First principles mechanistic simulations are a promising starting point to this end.

Over the longer-term, multiscale computational radiobiology will help usher in a new generation of virtual theranostic trials (VTTs) for nuclear medicine and radiation oncology. By integrating radiobiology with advanced PK frameworks (see companion paper in this special issue (*52*)) we aim to one day be able to routinely create digital twin avatars tailored to each patient. These personalized representations will serve as versatile, transformative platforms that allow for the sophisticated exploration of disease-specific and organ-level biological responses to RPTs in a controlled virtual environment. They promise to enable a variety of next-generation applications, including: patient-specific quality assurance informed by biological effectiveness; facilitated analysis of combination therapies through *in silico* testing of RPTs combined with external beam radiotherapy, chemotherapy, or immunotherapy; development and optimization of entirely new RPT clinical protocols; and adaptive optimization of therapies based on dynamic patient-specific data collection and model updating. In addition, VTTs will offer unique capabilities for evaluating the reproducibility and robustness of models across diverse cohorts and modalities, and serve as a powerful tool for education and training that allows clinicians, physicists, and trainees to explore complex scenarios in a risk-free environment.

These possibilities are tantalizing to imagine. At the same time, we must temper our enthusiasm and keep in mind that multiscale simulations are ultimately limited by our knowledge of the underlying mechanisms. As mentioned, although our collective understanding of the details of radiation response is relatively advanced compared to most other agents, RPT has highlighted that there are still open questions. Without answering these questions *qualitatively*, we have no chance of answering them *quantitatively*. Much work remains, and further progress will likely necessitate a combination of simulations, *in vitro* and preclinical experiments, and clinical translation.

In addition, it is important to realize that detailed mechanistic computations will probably never replace simpler effective models in the clinic. As we have repeatedly hinted at throughout this article, small models have many advantages over large ones that make them far more practical for day-to-day use, namely fewer parameters to fit to patient-specific data and drastically lower computational cost. At the same time, we must always remember the ever-present caveat that "All models are wrong, but some of them are useful." The trick is to know which simplifications are useful in a specific context. In our opinion, it is only through starting from a fundamental mechanistic description that we will be able to better understand the specific conditions in which the traditional LQ-based BED formalism is a reasonable approximation for RPT, and just as importantly (if not more so), the conditions where it is not, and how to modify it in a parsimonious way.

Realizing the promise of precision RPTs requires embracing a dynamic, multiscale view of radiobiology. It calls for the integration of experimental, clinical, and computational research to refine the definition of biological dose, capture biological effects more accurately, and ultimately optimize therapeutic outcomes. The future of RPT lies in embracing radiobiology as a living, evolving framework that questions deeply, models innovatively, and integrates data across biological scales to reveal the true therapeutic signatures of these uniquely powerful treatments. This is not merely an academic pursuit—it is a clinical imperative.

**References**

1. C. Joiner M, Van Der KogeJ. L A. Basic Clinical Radiobiology. 6th ed. Boca Raton: CRC Press; 2024.

2. Gardner LL, Thompson SJ, O'Connor JD, McMahon SJ. Modelling radiobiology. *Phys Med Biol*. 2024;69:18TR01.

3. Abdollahi H, Saboury B, Yusufaly T, Alberts I, Uribe C, Rahmim A. Shifting the Spotlight to Low Dose Rate Radiobiology in Radiopharmaceutical Therapies: Mathematical Modelling, Challenges and Future Directions. *IEEE Trans Radiat Plasma Med Sci*. 2025:1-1.

4. Terry SYA, Nonnekens J, Aerts A, et al. Call to arms: need for radiobiology in molecular radionuclide therapy. *Eur J Nucl Med Mol Imaging*. 2019;46:1588-1590.

5. Aerts A, Eberlein U, Holm S, et al. EANM position paper on the role of radiobiology in nuclear medicine. *Eur J Nucl Med Mol Imaging*. 2021;48:3365-3377.

6. Morris ZS, Wang AZ, Knox SJ. The Radiobiology of Radiopharmaceuticals. *Seminars in Radiation Oncology*. 2021;31:20-27.

7. McMahon SJ. The linear quadratic model: usage, interpretation and challenges. *Phys Med Biol*. 2018;64:01TR01.

8. Fowler JF. 21 years of Biologically Effective Dose. *BJR*. 2010;83:554-568.

9. Yusufaly T, Roncali E, Brosch-Lenz J, et al. Computational Nuclear Oncology Toward Precision Radiopharmaceutical Therapies: Current Tools, Techniques, and Uncharted Territories. *J Nucl Med*. 2025;66:509-515.

10. McMahon SJ, Prise KM. Mechanistic Modelling of Radiation Responses. *Cancers (Basel)*. 2019;11:205.

11. Lea DE, Catcheside DG. The mechanism of the induction by radiation of chromosome aberrations inTradescantia. *Journ of Genetics*. 1942;44:216-245.

12. Fowler JF. The linear-quadratic formula and progress in fractionated radiotherapy. *The British Journal of Radiology*. 1989;62:679-694.

13. Brenner DJ, Hlatky LR, Hahnfeldt PJ, Hall EJ, Sachs RK. A convenient extension of the linear-quadratic model to include redistribution and reoxygenation. *International Journal of Radiation Oncology*Biology*Physics*. 1995;32:379-390.

14. Zaider M, Hanin L. Tumor control probability in radiation treatment. *Medical Physics*. 2011;38:574-583.

15. Marks LB, Yorke ED, Jackson A, et al. Use of Normal Tissue Complication Probability Models in the Clinic. *International Journal of Radiation Oncology*Biology*Physics*. 2010;76:S10-S19.

16. Niemierko A. A generalized concept of equivalent uniform dose (EUD). *Med Phys*. 1999;26:1100.

17. Niemierko A. Reporting and analyzing dose distributions: A concept of equivalent uniform dose. *Medical Physics*. 1997;24:103-110.

18. Lyman JT. Complication Probability as Assessed from Dose-Volume Histograms. *Radiation Research*. 1985;104:S13.

19. Kutcher GJ, Burman C. Calculation of complication probability factors for non-uniform normal tissue irradiation: The effective volume method gerald. *International Journal of Radiation Oncology*Biology*Physics*. 1989;16:1623-1630.

20. Matsuya Y, Kai T, Sato T, et al. Track-structure modes in particle and heavy ion transport code system (PHITS): application to radiobiological research. *International Journal of Radiation Biology*. 2022;98:148-157.

21. Nikjoo S. Uehara W. E. Wilson M. Ho H. Track structure in radiation biology: theory and applications. *International Journal of Radiation Biology*. 1998;73:355-364.

22. Ortiz R, Ramos-Méndez J. A clustering tool for generating biological geometries for computational modeling in radiobiology. *Phys Med Biol*. 2024;69:21NT01.

23. Ward JF. The Complexity of DNA Damage: Relevance to Biological Consequences. *International Journal of Radiation Biology*. 1994;66:427-432.

24. Incerti S, Baldacchino G, Bernal M, et al. THE GEANT4-DNA PROJECT. *Int J Model Simul Sci Comput*. 2010;01:157-178.

25. Schuemann J, McNamara AL, Ramos-Méndez J, et al. TOPAS-nBio: An Extension to the TOPAS Simulation Toolkit for Cellular and Sub-cellular Radiobiology. *Radiation Research*. 2018;191:125.

26. Semenenko VA, Stewart RD. Fast Monte Carlo simulation of DNA damage formed by electrons and light ions. *Phys Med Biol*. 2006;51:1693-1706.

27. Tsai M-Y, Tian Z, Qin N, et al. A new open-source GPU-based microscopic Monte Carlo simulation tool for the calculations of DNA damages caused by ionizing radiation — Part I: Core algorithm and validation. *Medical Physics*. 2020;47:1958-1970.

28. Okada S, Murakami K, Incerti S, Amako K, Sasaki T. MPEXS-DNA, a new GPU-based Monte Carlo simulator for track structures and radiation chemistry at subcellular scale. *Med Phys*. 2019;46:1483-1500.

29. McMahon SJ, Prise KM. A Mechanistic DNA Repair and Survival Model (Medras): Applications to Intrinsic Radiosensitivity, Relative Biological Effectiveness and Dose-Rate. *Front Oncol*. 2021;11:689112.

30. Henthorn NT, Warmenhoven JW, Sotiropoulos M, et al. In Silico Non-Homologous End Joining Following Ion Induced DNA Double Strand Breaks Predicts That Repair Fidelity Depends on Break Density. *Sci Rep*. 2018;8:2654.

31. Folcik VA, An GC, Orosz CG. The Basic Immune Simulator: An agent-based model to study the interactions between innate and adaptive immunity. *Theor Biol Med Model*. 2007;4:39.

32. Walker DC, Hill G, Wood SM, Smallwood RH, Southgate J. Agent-Based Computational Modeling of Wounded Epithelial Cell Monolayers. *IEEE Trans.on Nanobioscience*. 2004;3:153-163.

33. Kunz LV, Bosque JJ, Nikmaneshi M, et al. AMBER: A Modular Model for Tumor Growth, Vasculature and Radiation Response. *Bull Math Biol*. 2024;86:139.

34. Gatenby RA, Brown JS. Integrating evolutionary dynamics into cancer therapy. *Nat Rev Clin Oncol*. 2020;17:675-686.

35. Cai Y, Zhang J, Li Z. Multi-scale mathematical modelling of tumour growth and microenvironments in anti-angiogenic therapy. *BioMed Eng OnLine*. 2016;15:155.

36. An G, Mi Q, Dutta-Moscato J, Vodovotz Y. Agent-based models in translational systems biology. *WIREs Mechanisms of Disease*. 2009;1:159-171.

37. Liu R, Higley KA, Swat MH, Chaplain MAJ, Powathil GG, Glazier JA. Development of a coupled simulation toolkit for computational radiation biology based on Geant4 and CompuCell3D. *Phys Med Biol*. 2021;66:045026.

38. García García OR, Ortiz R, Moreno-Barbosa E, D-Kondo N, Faddegon B, Ramos-Méndez J. TOPAS-Tissue: A Framework for the Simulation of the Biological Response to Ionizing Radiation at the Multi-Cellular Level. *International Journal of Molecular Sciences*. 2024;25:10061.

39. Sgouros G, Hobbs RF. Dosimetry for Radiopharmaceutical Therapy. *Seminars in Nuclear Medicine*. 2014;44:172-178.

40. Strand S, Zanzonico P, Johnson TK. Pharmacokinetic modeling. *Medical Physics*. 1993;20:515-527.

41. Zanzonico P. The MIRD Schema for Radiopharmaceutical Dosimetry: A Review. *J Nucl Med Technol*. 2024;52:74-85.

42. Imperiale A, Jha A, Meuter L, Nicolas GP, Taïeb D, Pacak K. The Emergence of Somatostatin Antagonist–Based Theranostics: Paving the Road Toward Another Success? *J Nucl Med*. 2023;64:682-684.

43. Tamborino G, Engbers P, De Wolf TH, et al. Establishing In Vitro Dosimetric Models and Dose–Effect Relationships for $^{177}$Lu-DOTATATE in Neuroendocrine Tumors. *J Nucl Med*. May 2025:jnumed.125.269470.

44. Lim A, Andriotty M, Yusufaly T, Agasthya G, Lee B, Wang C. A fast Monte Carlo cell-by-cell simulation for radiobiological effects in targeted radionuclide therapy using pre-calculated single-particle track standard DNA damage data. *Front Nucl Med*. 2023;3:1284558.

45. Tamborino G, De Saint-Hubert M, Struelens L, et al. Cellular dosimetry of [177Lu]Lu-DOTA-[Tyr3]octreotate radionuclide therapy: the impact of modeling assumptions on the correlation with in vitro cytotoxicity. *EJNMMI Phys*. 2020;7:8.

46. Golshani M, Mowlavi AA, Azadegan B. Computational assessment of the cellular dosimetry and microdosimetry of the gadolinium electrons released during neutron capture therapy. *Biomed Phys Eng Express*. 2019;5:025031.

47. Bastiaannet R, Liatsou I, F Hobbs R, Sgouros G. Large-scale in vitro microdosimetry via live cell microscopy imaging: implications for radiosensitivity and RBE evaluations in alpha-emitter radiopharmaceutical therapy. *J Transl Med*. 2023;21:144.

48. Katugampola S, Hobbs RF, Howell RW. Generalized methods for predicting biological response to mixed radiation types and calculating equieffective doses (EQDX). *Medical Physics*. 2024;51:637-649.

49. Peter R, Bidkar AP, Bobba KN, et al. 3D small-scale dosimetry and tumor control of 225Ac radiopharmaceuticals for prostate cancer. *Sci Rep*. 2024;14:19938.

50. Li WB, Bouvier-Capely C, Saldarriaga Vargas C, Andersson M, Madas B. Heterogeneity of dose distribution in normal tissues in case of radiopharmaceutical therapy with alpha-emitting radionuclides. *Radiat Environ Biophys*. 2022;61:579-596.

51. Kratochwil C, Bruchertseifer F, Rathke H, et al. Targeted α-Therapy of Metastatic Castration-Resistant Prostate Cancer with $^{225}$Ac-PSMA-617: Dosimetry Estimate and Empiric Dose Finding. *J Nucl Med*. 2017;58:1624-1631.

52. Hardiansyah D et al. An overview of PBPK and PopPK Models: Applications to Radiopharmaceutical Therapies for Analysis and Personalization. *PET Clinics*. 2026.